\documentclass[aps,11pt,pra,showpacs,superscriptaddress,preprint]{revtex4}

\usepackage{amsmath}
\usepackage{graphicx}
\usepackage{subfig}
\usepackage{array}

\catcode`ð=\active
\defð{\u{g}}
\catcode`Ð=\active
\defÐ{\u{G}}
\catcode`Ý=\active
\defÝ{\. I}
\catcode`ö=\active
\defö{\"{o}}
\catcode`Ö=\active
\defÖ{\"O}
\catcode`ü=\active
\defü{\"{u}}
\catcode`Ü=\active
\defÜ{\"{U}}
\catcode`Þ=\active
\defÞ{\c{S}}
\catcode`þ=\active
\defþ{\c{s}}
\catcode`ý=\active
\defý{{\i}}
\catcode`ç=\active
\defç{\d{c}}
\catcode`Ç=\active
\defÇ{\d{C}}

\begin{document}

\title{Effective-Mass Klein-Gordon-Yukawa Problem for Bound and Scattering States}
\author{\small Altuð Arda}
\email[E-mail: ]{arda@hacettepe.edu.tr}\affiliation{Department of
Physics Education, Hacettepe University, 06800, Ankara,Turkey}
\author{\small Ramazan Sever}
\email[E-mail: ]{sever@metu.edu.tr}\affiliation{Department of
Physics, Middle East Technical  University, 06531, Ankara,Turkey}

\begin{abstract}

Bound and scattering state solutions of the effective-mass
Klein-Gordon equation are obtained for the Yukawa potential with
any angular momentum $\ell$. Energy eigenvalues, normalized wave
functions and scattering phase shifts are calculated as well as
for the constant mass case. Bound state solutions of the Coulomb
potential are also studied as a limiting
case. Analytical and numerical results are compared with the ones obtained before.\\
Keywords: Yukawa potential, Coulomb potential, Klein-Gordon
equation, Position-Dependent Mass, Bound State, Scattering State
\end{abstract}
\pacs{03.65N, 03.65.Ge, 03.65.Nk, 03.65.Pm, 03.65.-w, 12.39.Fd}

\maketitle

\newpage

\section{Introduction}

In the view of relativistic quantum mechanics, a particle moving
in a potential field is described particularly with the
Klein-Gordon (KG) equation. Solutions of the  one-dimensional KG
equation have been received great attention for some potentials
[1-2]. The relativistic quantum mechanical problems that can be
solved exactly are very restricted [3]. In the present work, we
obtain approximate analytical energy eigenvalues, normalized wave
functions and scattering phase shifts for the Yukawa potential [4]
\begin{eqnarray}
V(r)=-\frac{\eta}{r}\,e^{-\,\alpha r}\,.
\end{eqnarray}
where $\alpha$ is the screening parameter and $\eta$ is the
strength of the potential. The Yukawa potential has many
applications in different areas of physics: in high-energy physics
as a short-range potential [4], atomic and molecular physics as a
screened Coulomb potential and plasma physics as the Debye-Hückel
potential [5]. In recent years, considerable efforts have also
been made to study the approximate bound state solutions of the
Yukawa potential in the non-relativistic domain by using different
methods [5-9].

On the other hand, the position-dependent mass (PDM) formalism
[10] has many applications in different areas, such as impurities
in crystals [11], the study of quantum wells and quantum dots [12]
and semiconductor heterostructures [13]. In recent years, the
relativistic extension of the position-dependent mass formalism
has been studied by many authors for different types of potentials
[14-16].

The organization of this work is as follows. In Section II, we
study the approximate bound state solutions and corresponding
normalized wave functions for the Yukawa potential. We list some
numerical results for the cases of PDM and constant mass presented
in Table I and II. In Section III, we deal with the approximate
scattering state solutions of the Yukawa potential and give
analytical expressions for the phase shifts. In Section IV, we
give our conclusions.

\section{Bound State Solutions}
The radial part of the effective-mass KG equation is written as
[17]
\begin{eqnarray}
\frac{d^2\phi(r)}{dr^2}-\bigg\{\frac{\ell(\ell+1)}{r^2}+\frac{1}{\hbar^2
c^2}\big[m^{2}(r)c^4-(E^2-2EV(r)+V^2(r))\big]\bigg\}\phi(r)=0\,,
\end{eqnarray}
where $\ell$ is the angular momentum quantum number, $E$ is the
energy of the particle and $c$ is the velocity of the light.

In recent years, the following approximation
\begin{eqnarray}
\frac{1}{r^2}\approx4\alpha^2\frac{e^{-2\alpha r}}{(1-e^{-2\alpha
r})^2}\,,
\end{eqnarray}
has been used [18, 19] instead of the centrifugal term in the wave
equations to obtain the solutions with any $\ell$ values. It has a
good accuracy for small values of the potential parameter $\alpha$
[18, 19]. A remarkable approximation is proposed by Alhaidari [20]
where the author suggested, for the first time, an approximation
for the orbital term $1/r$ in the Dirac equation not for the
$1/r^2$ term. By using this approximation, it is possible to find
the approximate bound state solutions of the Dirac equation for
coupling to pure $1/r$ vector potentials with any $\ell$ values
for higher excitation levels with more accuracy than using the
traditional approximation for the $1/r^2$ term [20].

We define the mass function as
\begin{eqnarray}
m(r)=m_0+\frac{m_1}{e^{2\alpha r}-1}\,,
\end{eqnarray}
where $m_0$ and $m_1$ are two parameters and $m_0$ will correspond
to the rest mass of the KG particle. Using the approximation given
in Eq. (3) Yukawa potential becomes
\begin{eqnarray}
V(r)=-2\alpha\eta\,\frac{e^{-2\alpha r}}{1-e^{-2\alpha r}}\,,
\end{eqnarray}
Inserting Eqs. (4) and (5) into Eq. (2) and taking a new variable
$z=(1-e^{-2\alpha r})^{-1}$ ($z \rightarrow \infty$ for $r
\rightarrow 0$ and $z \rightarrow 1$ for $r \rightarrow \infty$),
we obtain
\begin{eqnarray}
&&z(1-z)\frac{d^2\phi(z)}{dz^2}+(1-2z)\frac{d\phi(z)}{dz}\nonumber\\&+&
\bigg\{\ell(\ell+1)-\frac{\beta^2}{4\alpha^2}\left(m_0^2c^4-E^2\right)\frac{1}{z(1-z)}
+\left(\frac{\beta^2m_0m_1c^4}{2\alpha^2}-\frac{E\beta^2\eta}{2\alpha}\right)\frac{1}{z}\nonumber\\&-&
\left(\frac{\beta^2m_1^2c^4}{4\alpha^2}-\beta^2\eta^2\right)\frac{1-z}{z}\bigg\}\phi(z)=0\,,
\end{eqnarray}
where $\beta^2=1/\hbar^2c^2$. Taking the form of the wave function
\begin{eqnarray}
\phi(z)=z^{\lambda_{1}}(1-z)^{\lambda_{2}}\psi(z)\,,
\end{eqnarray}
and inserting into Eq. (6), one  gets a hypergeometric-type
equation [21]
\begin{eqnarray}
&&z(1-z)\frac{d^2\psi(z)}{dz^2}+[1+2\lambda_{1}-2(\lambda_{1}+\lambda_{2}+1)z]\frac{d\psi(z)}{dz}\nonumber\\&+&
\big\{-\lambda^2_{1}-\lambda^2_{2}-\lambda_{1}-\lambda_{2}-2\lambda_{1}\lambda_{2}
+\ell(\ell+1)+\frac{\beta^2m_1^2c^4}{4\alpha^2}-\beta^2\eta^2\big\}\psi(z)=0\,,
\end{eqnarray}
where
\begin{eqnarray}
\lambda^2_{1}&=&\frac{\beta^2}{4\alpha^2}\left(m_0^2c^4-E^2\right)-\frac{\beta^2m_0m_1c^4}{2\alpha^2}
+\frac{\beta^2m_1^2c^4}{4\alpha^2}+\frac{E\beta^2\eta}{2\alpha}-\beta^2\eta^2\,,\\
\lambda^2_{2}&=&\frac{\beta^2}{4\alpha^2}\left(m_0^2c^4-E^2\right)\,.
\end{eqnarray}

Comparing Eq. (8) with the hypergeometric equation of the
following form  [21]
\begin{eqnarray}
z(1-z)y''+[\xi_3-(\xi_1+\xi_2+1)z]y'-\xi_1\xi_2y=0\,,
\end{eqnarray}
we find the solution of Eq. (8) as the hypergeometric function
\begin{eqnarray}
\psi(z)=\,_2F_1(\xi_1,\xi_2;\xi_3;z)\,.
\end{eqnarray}
where
\begin{eqnarray}
\xi_1&=&\lambda_{1}+\lambda_{2}+\frac{1}{2}\bigg(1+\sqrt{1+4\ell(\ell+1)
+\frac{\beta^2m_1^2c^4}{\alpha^2}-4\beta^2\eta^2\,}\,\bigg)\,,\\
\xi_2&=&\lambda_{1}+\lambda_{2}+\frac{1}{2}\bigg(1-\sqrt{1+4\ell(\ell+1)
+\frac{\beta^2m_1^2c^4}{\alpha^2}-4\beta^2\eta^2\,}\,\bigg)\,,\\
\xi_3&=&1+2\lambda_{1}\,.
\end{eqnarray}
From Eq. (7), we obtain total wave function
\begin{eqnarray}
\phi(z)=Nz^{\lambda_{1}}(1-z)^{\lambda_{2}}\,_2F_1(\xi_1,\xi_2,\xi_3,z)\,.
\end{eqnarray}
where $N$ is normalization constant that will be derived in
Appendix A. When either $\xi_1$ or $\xi_2$ equals to a negative
integer $-n$, the hypergeometric function $\psi(z)$ can be reduced
to a finite solution. This gives us a polynomial of degree $n$ in
Eq. (12) and the following quantum condition
\begin{eqnarray}
\lambda_{1}+\lambda_{2}+\frac{1}{2}+\frac{1}{2}\sqrt{1+4\ell(\ell+1)
+\frac{\beta^2m_1^2c^4}{\alpha^2}-4\beta^2\eta^2\,}=-n\,\,\,(n=0,
1, 2, \ldots).
\end{eqnarray}

It is the relativistic energy eigenvalue equation for the Yukawa
potential within the PDM formalism. Defining two new parameters
such as
\begin{eqnarray}
\mathcal{L}(\ell)&=&\sqrt{1+4\ell(\ell+1)+\frac{\beta^2m_1^2c^4}{\alpha^2}-4\beta^2\eta^2\,}\,,\\
\lambda^2_{1}&=&\frac{\beta^2}{4\alpha^2}\left(m_0^2c^4-E^2\right)+\frac{E\beta^2\eta}{2\alpha}+\Lambda(\beta)\,;\,\,
\Lambda(\beta)=-\frac{\beta^2m_0m_1c^4}{2\alpha^2}+\frac{\beta^2m_1^2c^4}{4\alpha^2}-\beta^2\eta^2\,,
\end{eqnarray}
we get the approximate energy eigenvalues as
\begin{eqnarray}
&&E^{(\mp)}_{n\ell}=\frac{\alpha\eta}{2[(\mathcal{L}(\ell)+2n+1)^2+\beta^2\eta^2]}
\bigg\{[\mathcal{L}(\ell)+2n+1]^2-4\Lambda(\beta)
\nonumber\\&\mp&\frac{\mathcal{L}(\ell)+2n+1}{\beta\eta}\nonumber\\&\times&
\sqrt{\frac{4\beta^2m_0^2}{\alpha^2}\,\left[\mathcal{L}(\ell)+2n+1\right]^2+\beta^2\eta^2\big]
-\big[[\mathcal{L}(\ell)+2n+1]^2-4\Lambda(\beta)\big]^2\,}\bigg\}\,.
\end{eqnarray}

Table I presents the comparison of our numerical results for the
case of constant mass ($m_1=0$) with the ones given in Ref. [22].
We restrict ourselves for only $s$-states and take $m_0=1$ because
of the computation in Ref. [22]. Our parameters $\eta$ and
$\alpha$ correspond to $\lambda$ and $k(\equiv \omega\lambda)$,
respectively. The relativistic energy is obtained as
$E_R=\eta_{exact}\left(\sqrt{1-\lambda^2\,}-1\right)+1$ in Ref.
[22]. The same numerical values of $\eta_{exact}$ is used with
Ref. [22] to compare our numerical results. We plot the $1/r$ and
the approximation $2\alpha e^{-\alpha r }(1-e^{-\alpha r })$
versus $r$. It seems that the energy eigenvalues have a good
accuracy up to the values of $\eta\leq0.25$ and $\alpha\leq0.30$.
Table II presents numerical energy eigenvalues for the case of
position dependent mass including also for the case of constant
mass with different values of ($n, \ell$).

Now, let us study the results of our formalism for the case of the
Coulomb potential.

\subsubsection{Relativistic-Coulomb Limit}

In the limiting case $\alpha \rightarrow 0$, the Yukawa potential
reduces to
\begin{eqnarray}
V(r)=-\frac{\eta}{r}\,,
\end{eqnarray}
which is an attractive Coulomb potential received great interest
not only in the case of constant mass [23-25] but also within the
position-dependent mass formalism [26].

In order to compare our results for the bound states with the ones
obtained in Ref. [27], we expand the mass function, Eq. (4), into
Taylor series
\begin{eqnarray}
m(r) \xrightarrow[\alpha \to 0]{}
m_0-\frac{m_1}{2}+\frac{m_1}{2\alpha r}+\ldots\equiv
M_0+\frac{M_1}{r}+\ldots\,,
\end{eqnarray}

The parameters $b$ and $m_0$ used in Ref. [27] are defined as
$b\rightarrow M_1, m_0 \rightarrow M_0$. Taking the vector part of
the potential which is equal to the scalar part as stated in Ref.
[27] as $V(r)=S(r)=-\frac{\eta}{r}$ and inserting the mass
function into the KG equation including the scalar potential
\begin{eqnarray}
\frac{d^2\phi(r)}{dr^2}-\bigg\{\frac{\ell(\ell+1)}{r^2}+\frac{1}{\hbar^2
c^2}\big[m^{2}(r)c^4+2m(r)S(r)+S^2(r)-(E^2-2EV(r)+V^2(r))\big]\bigg\}\phi(r)=0\,,\nonumber\\
\end{eqnarray}
gives the following equation
\begin{eqnarray}
\left\{\frac{d^2}{dr^2}-A^2_1-A_2\,\frac{1}{r}-A_3\frac{1}{r^2}\right\}\phi(r)=0\,,
\end{eqnarray}
where
\begin{eqnarray}
A^2_1=\beta^2(M^2_0c^4-E^2)\,;A_2=\beta^2(2M_0M_1c^4-2M_0c^2\eta-2E\eta)
\,;A_3=\beta^2(M^2_1c^4-2M_1c^2\eta+\beta\ell(\ell+1))\,,\nonumber\\
\end{eqnarray}
The wave function is written as
\begin{eqnarray}
\phi(r)=r^{\kappa}e^{-A_{1}r}f(r)\,.
\end{eqnarray}
where we set $\kappa(\kappa-1)=A_3$. Thus, inserting Eq. (26) into
Eq. (24) and using a new transformation $z=2A_{1}r$, we obtain
\begin{eqnarray}
z\frac{d^2f(z)}{dz^2}+(2\kappa-z)\frac{df(z)}{dz}+(-\kappa-\frac{A_2}{2A_{1}})f(z)=0\,.
\end{eqnarray}
which has the form of the Kummer differential equation [21]
\begin{eqnarray}
xy''(x)+(c-x)y'(x)-ay(x)=0\,.
\end{eqnarray}
So, the solution of Eq. (27) is given by
\begin{eqnarray}
f(z) \sim\,_1F_1\left(\kappa+\frac{A_2}{2A_1};2\kappa,z\right)\,.
\end{eqnarray}
In order to get a finite solution, the following condition must be
satisfied
\begin{eqnarray}
\kappa+\frac{A_2}{2A_1}=-n\,\,\,(n=0, 1, 2, \ldots)\,.
\end{eqnarray}
We get the bound state energy eigenvalues for the Coulomb
potential as
\begin{eqnarray}
E^{Coul.}_{n\ell}=\frac{M_0
c^2}{4\beta^2\eta^2+\left[N+\sqrt{1+\eta'+4\beta^2\ell(\ell+1)\,}\,\right]^2}
\times\bigg\{4\beta^2\eta(M_1c^2-\eta)\nonumber\\+\left(N+\sqrt{1+\eta'+4\beta^2\ell(\ell+1)\,}
\right)\sqrt{\left[N+\sqrt{1+\eta'+4\beta^2\ell(\ell+1)\,}\,\right]^2-\eta'(M_1c^2-2\eta)\,}\,\bigg\}\,.
\end{eqnarray}
where
\begin{eqnarray}
N=2n+1\,\,\, \text{and}\,\,\, \eta'=4\beta
M_{1}c^2\left(M_{1}c^2-2\eta\right)\,.
\end{eqnarray}
This result is the same with the one for $\ell=0$ given in Ref.
[27].

\section{Scattering State Solutions}
Now we turn to the solution of the Eq. (2) to obtain the
scattering states for the Yukawa potential. We use a new variable
$s=1-e^{-2\alpha r }$ ($s \rightarrow 0$ for $r \rightarrow 0$ and
$s \rightarrow 1$ for $r \rightarrow \infty$) and obtain
\begin{eqnarray}
&&s(1-s)\frac{d^2\phi(s)}{ds^2}-s\frac{d\phi(s)}{ds}\nonumber\\&+&
\bigg\{\frac{E\beta^2\eta}{2\alpha}-\frac{\beta^2}{2\alpha^2}\left(m_0m_1c^4-\ell(\ell+1)\right)\frac{1}{s}
-\frac{\beta^2}{4\alpha^2}\left(m_0^2c^4-E^2\right) \frac{s}{1-s}
\nonumber\\&+&
\left(\beta^2\eta^2-\frac{\beta^2m_1^2c^4}{4\alpha^2}\right)\frac{1-s}{s}\bigg\}\phi(s)=0\,.
\end{eqnarray}

Defining the trial wave function
\begin{eqnarray}
\phi(s)=s^{k_1}(1-s)^{k_2}\psi(s)\,,
\end{eqnarray}
and substituting into Eq. (33), we obtain a hypergeometric-type
equation for $\psi(s)$
\begin{eqnarray}
&&s(1-s)\frac{d^2\psi(s)}{ds^2}+[2k_1-(2k_1+2ik'_2+1)s]\frac{d\psi(s)}{ds}\nonumber\\&+&
\big\{-2ik_1k'_2-k_1-\ell(\ell+1)-\frac{\beta^2m_1^2c^4}{2\alpha^2}+\frac{E\beta^2\eta}{2\alpha}\big\}\psi(s)=0\,,
\end{eqnarray}
where
\begin{eqnarray}
k_1&=&\frac{1}{2}\bigg\{1+\sqrt{1+4\ell(\ell+1)+\frac{\beta^2m^2_1c^4}{\alpha^2}-4\beta^2\eta^2\,}\bigg\}\,,\\
k_2&=&ik'_2\,\,\,;\,k'_2=\sqrt{\frac{\beta^2}{4\alpha^2}\left(E^2-m_0^2c^4\right)\,}\,.
\end{eqnarray}

The solution of Eq. (35) is a hypergeometric function
\begin{eqnarray}
\psi(s)=\,_2F_1(p,q;r;s)\,,
\end{eqnarray}
where
\begin{eqnarray}
p&=&k_1+ik'_2+\sqrt{\frac{\beta^2m_1^2c^4}{4\alpha^2}-\beta^2\eta^2-\frac{\beta^2}{4\alpha^2}\left(E^2-m_0^2c^4\right)-
\frac{\beta^2m_0m_1c^4}{2\alpha^2}+\frac{E\beta^2\eta}{2\alpha}\,}\,,\\
q&=&k_1+ik'_2-\sqrt{\frac{\beta^2m_1^2c^4}{4\alpha^2}-\beta^2\eta^2-\frac{\beta^2}{4\alpha^2}\left(E^2-m_0^2c^4\right)-
\frac{\beta^2m_0m_1c^4}{2\alpha^2}+\frac{E\beta^2\eta}{2\alpha}\,}\,,\\
r&=&2k_1\,.
\end{eqnarray}
From Eqs. (34) and (38), we write the wave function of the
scattering states
\begin{eqnarray}
\phi(s)=s^{k_1}(1-s)^{ik'_2}\,_2F_1(p,q;r;s)\,,
\end{eqnarray}
or
\begin{eqnarray}
\phi(r)=(1-e^{-2\alpha r})^{k_1}e^{-2ik'_2\alpha r
}\,_2F_1(p,q;r;1-e^{-2\alpha r})\,.
\end{eqnarray}

To obtain a finite solution, $p$ or $q$ must be a negative
integer. This gives the following equality
\begin{eqnarray}
k_1+ik'_2+\sqrt{\frac{\beta^2m_1^2c^4}{4\alpha^2}-\beta^2\eta^2-\frac{\beta^2}{4\alpha^2}\left(E^2-m_0^2c^4\right)-
\frac{\beta^2m_0m_1c^4}{2\alpha^2}+\frac{E\beta^2\eta}{2\alpha}\,}=-n\,,\,(n=0,
1, 2,\ldots)\nonumber\\
\end{eqnarray}
which is the same energy eigenvalue equation given in Eq. (17). We
write the asymptotic form of the wave function given in Eq. (43)
to obtain the scattering phase shifts. For this purpose, we use
the property of the hypergeometric functions [21]
\begin{eqnarray}
_2F_1(a,b;c;y)&=&\frac{\Gamma(c)\Gamma(c-a-b)}{\Gamma(c-a)\Gamma(c-b)}\,_2F_1(a,b;a+b-c+1;1-y)\nonumber\\&+&
\frac{\Gamma(c)\Gamma(a+b-c)}{\Gamma(a)\Gamma(b)}\,(1-y)^{c-a-b}\,_2F_1(c-a,c-b;c-a-b+1;1-y)\,,
\end{eqnarray}
and $_2F_1(a,b;c;0)=1$, we obtain the wave function for the limit
of $r \rightarrow \infty$
\begin{eqnarray}
\phi(r \rightarrow \infty)\rightarrow(1&-&e^{-2\alpha
r})^{k_1}\bigg\{\frac{\Gamma(2k_1)\Gamma(-2ik'_2)}{\Gamma(k_1-ik'_2-\mathcal{A}(k_1,k_2))
\Gamma(k_1-ik'_2+\mathcal{A}(k_1,k_2))}\,e^{-2ik'_2\alpha
r}\nonumber\\&+&\frac{\Gamma(2k_1)\Gamma(2ik'_2)}{\Gamma(k_1+ik'_2-\mathcal{A}(k_1,k_2))
\Gamma(k_1+ik'_2+\mathcal{A}(k_1,k_2))}\,e^{2ik'_2\alpha
r}\bigg\}\,,
\end{eqnarray}
which could be written
\begin{eqnarray}
\phi(r \rightarrow \infty)\rightarrow(1&-&e^{-2\alpha
r})^{k_1}\Gamma(2k_1)\bigg\{\left[\frac{\Gamma(2ik'_2)}{\Gamma(k_1+ik'_2-\mathcal{A}(k_1,k_2))
\Gamma(k_1+ik'_2+\mathcal{A}(k_1,k_2))}\,\right]^{*}e^{-2ik'_2\alpha
r}\nonumber\\&+&\frac{\Gamma(2ik'_2)}{\Gamma(k_1+ik'_2-\mathcal{A}(k_1,k_2))
\Gamma(k_1+ik'_2+\mathcal{A}(k_1,k_2))}\,e^{2ik'_2\alpha
r}\bigg\}\,,
\end{eqnarray}
where
\begin{eqnarray}
\mathcal{A}(k_1,k_2)=\sqrt{\frac{\beta^2m_1^2c^4}{4\alpha^2}-\beta^2\eta^2-\frac{\beta^2}{4\alpha^2}
\left(E^2-m_0^2c^4\right)-
\frac{\beta^2m_0m_1c^4}{2\alpha^2}+\frac{E\beta^2\eta}{2\alpha}\,}\,.
\end{eqnarray}
From Eq. (47) we obtain
\begin{eqnarray}
&&\phi(r \rightarrow \infty)\rightarrow 2(1-e^{-2\alpha
r})^{k_1}\Gamma(2k_1)\left|\frac{\Gamma(2ik'_2)}{\Gamma(k_1+ik'_2-\mathcal{A}(k_1,k_2))
\Gamma(k_1+ik'_2+\mathcal{A}(k_1,k_2))}\right|\nonumber\\&&
sin\left(2\alpha k'_2
r+\frac{\pi}{2}+arg\frac{\Gamma(2ik'_2)}{\Gamma(k_1+ik'_2-\mathcal{A}(k_1,k_2))
\Gamma(k_1+ik'_2+\mathcal{A}(k_1,k_2))}\right)\,,
\end{eqnarray}
and, consequently, the phase shifts $\delta_{\ell}$ are obtained
as
\begin{eqnarray}
\delta_{\ell}=(\ell+1)\frac{\pi}{2}+\delta=(\ell+1)\frac{\pi}{2}+
arg\frac{\Gamma(2ik'_2)}{\Gamma(k_1+ik'_2-\mathcal{A}(k_1,k_2))
\Gamma(k_1+ik'_2+\mathcal{A}(k_1,k_2))}\,.
\end{eqnarray}

\section{Conclusion}

We have studied the approximate bound and scattering state
solutions of the effective mass KG equation for the Yukawa
potential. We have obtained the energy eigenvalues, normalized
wave functions and scattering phase shifts approximately as well
as for the constant mass case. We have presented our numerical
results of the energy eigenvalues in Tables I and II. We have
compared them for the constant mass case with the ones obtained in
the literature. We have also studied the bound state solutions of
the Coulomb potential obtained from the limiting case of $\alpha
\rightarrow 0$ with the position-dependent and constant mass
cases. We have observed that the results obtained for the Coulomb
potential are the same with the ones obtained in the literature.

\section{Acknowledgments}
This research was partially supported by the Scientific and
Technical Research Council of Turkey.

\appendix

\section{Derivation of Normalization Constant}
By using Eq. (17), the wavefunction in Eq. (16) is written
\begin{eqnarray}
\phi(z)=Nz^{\lambda_{1}}(1-z)^{\lambda_{2}}\,_2F_1(-n,n+2\lambda_{1}+2\lambda_{2}+1;1+2\lambda_{1};z)\,,
\end{eqnarray}
We use a new variable $s=z^{-1}$ ($s \rightarrow 0$ for $z
\rightarrow \infty$ and $s \rightarrow 1$ for $z \rightarrow 1$)
to normalize the wavefunction. For this purpose, we substitute $s
\rightarrow \frac{2}{1-s}$
\begin{eqnarray}
\phi\big(\frac{1-s}{2}\big)=N'(1-s)^{\lambda_{1}}(1+s)^{\lambda_{2}}\,_2F_1(-n,n+2\lambda_{1}
+2\lambda_{2}+1;1+2\lambda_{1};\frac{1-s}{2})\,,
\end{eqnarray}
where $N'=N2^{-(\lambda_{1}+\lambda_{2})}(-1)^{2\lambda_{2}}$\,.

Using the following equality [21]
\begin{eqnarray}
P^{(\alpha,\beta)}_{n}(x)=\frac{(1+\alpha)_{n}}{n!}\,_2F_1(-n,n+\alpha+\beta+1;1+\alpha;\frac{1-x}{2})\,,
\end{eqnarray}
the hypergeometric functions in Eq. (A2) could be written in terms
of the Jacobi polynomials $P^{(\alpha,\beta)}_{n}(x)$. Here,
$(\kappa)_n$ is defined
$(\kappa)_n=\frac{\Gamma(\kappa+n)}{\Gamma(\kappa)}$, and setting
$\alpha=2\lambda_{1}$ and $\beta=2\lambda_{2}$ in Eq. (A3), we
rewrite Eq. (A2)
\begin{eqnarray}
\phi\big(\frac{1-s}{2}\big)=N'(1-s)^{\lambda_{1}}(1+s)^{\lambda_{2}}
n!\frac{\Gamma(1+2\lambda_{1})}{\Gamma(n+1+2\lambda_{1})}
P^{(2\lambda_{1},2\lambda_{2})}_{n}(s)\,.
\end{eqnarray}
where the Jacobi polynomials $P^{(a,b)}_{n}(z)$ are defined [21]
\begin{eqnarray}
P^{(a,b)}_{n}(x)=\frac{1}{n!}\sum_{k=0}^{n}\frac{\Gamma(-n+k)}{\Gamma(-n)}\frac{\Gamma(a+b+n+k+1)}{\Gamma(a+b+n+1)}
\frac{\Gamma(a+n+1)}{\Gamma(a+k+1)}\frac{1}{k!}(1-x)^{k}2^{-k}\,,
\end{eqnarray}

Using Eq. (A5) in Eq. (A4) we obtain the wavefunction
\begin{eqnarray}
\phi\big(\frac{1-s}{2}\big)=N'\Sigma(n,k)(1-s)^{\lambda_{1}+k}(1+s)^{\lambda_{2}}\,,
\end{eqnarray}
where
\begin{eqnarray}
\Sigma(n,k)=\Gamma(1+2\lambda_{1})\sum_{k=0}^{n}\frac{1}{2^{k}k!}\frac{\Gamma(-n+k)}{\Gamma(-n)}
\frac{\Gamma(2\lambda_{1}+2\lambda_{2}+n+k+1)}{\Gamma(2\lambda_{1}+2\lambda_{2}+n+1)}
\frac{1}{\Gamma(2\lambda_{1}+k+1)}\,.
\end{eqnarray}

The normalization condition
$\int_{0}^{1}\left|\phi\big(\frac{1-s}{2}\big)\right|^2ds=1$ gives
\begin{eqnarray}
\left|N'\right|^2\left|\Sigma(n,k)\right|^2\int_{0}^{1}(1-s)^{2\lambda_{1}+2k}(1+s)^{2\lambda_{2}}ds=1\,.
\end{eqnarray}
Comparing the last integral with the following [21]
\begin{eqnarray}\int_{0}^{1}t^{\delta-1}(1-t)^{\nu-\delta-1}(1-zt)^{-\gamma}dt=\frac{\Gamma(\delta)\Gamma(\nu-\delta)}
{\Gamma(\nu)}\,_2F_1(\gamma,\delta;\nu;z)\,,
\end{eqnarray}
and setting $\delta=1$, $z=-1$, $\gamma=-2\lambda_{2}$ and
$\nu=2\lambda_{1}+2k+2$ we find the normalization constant as
\begin{eqnarray}
\left|N'\right|^2=\frac{2\lambda_{1}+2k+1}{\left|\Sigma(n,k)\right|^2\,_2F_1(-2\lambda_{2},1;2\lambda_{1}+2k+2;-1)}\,.
\end{eqnarray}

\newpage

\begin{table}
\caption{Comparison of Klein-Gordon ground state energies for
different parameter values.}
\begin{ruledtabular}
\begin{tabular}{ccccc}
$\eta$ & $\alpha$ & $\eta_{exact}$ & Ref. [22] & our results\\
0.125 & 0.01250 & 0.83072460 & 0.993484 & 0.998702 \\
 & 0.06250 & 0.30947218 & 0.997573 & 0.999999 \\
 & 0.09375 & 0.12370738 & 0.999030 & 0.999542 \\
 & 0.12500 & 0.02452195 & 0.999808 & 0.998100 \\
 & 0.14375 & 0.00187260 & 0.999985 & 0.996759 \\
0.25 & 0.0250 & 0.88881431 & 0.971776 & 0.994130 \\
 & 0.1250 & 0.35655334 & 0.998678 & 0.999960 \\
 & 0.1875 & 0.15650395 & 0.995030 & 0.998556 \\
 & 0.2500 & 0.04068600 & 0.998708 & 0.993138 \\
 & 0.3000 & 0.00257940 & 0.999918 & 0.985799 \\
\end{tabular}
\end{ruledtabular}
\end{table}

\newpage

\begin{table}
\begin{ruledtabular}
\caption{Energy eigenvalues of the Yukawa potential for different
values of $n$ and $\ell$ (in $\hbar=m_0=c=1$ unit).}
\begin{tabular}{@{}ccllcccc@{}}
&&&&\multicolumn{2}{c}{$E^{(+)}_{n\ell}$}
&\multicolumn{2}{c}{$-E^{(-)}_{n\ell}$} \\ \cline{5-6} \cline{7-8}
$n$ & $\ell$ & $\eta$ & $\alpha$ & $m_{1}=0$ & $m_{1}=0.1$ & $m_{1}=0$ & $m_{1}=0.1$   \\
0 & 0 & 0.1 & 0.01 & 0.999181 & 0.411464 & 0.998173 & 0.394898 \\
 &  & 0.01 & 0.1 & 0.995475 & 0.859773 & 0.994475 & 0.855513 \\
1 & 0 & 0.1 & 0.01 & 0.999987 & 0.614868 & 0.998985 & 0.602709 \\
 &  & 0.01 & 0.1 & 0.980294 & 0.900967 & 0.979294 & 0.898992 \\
 & 1 & 0.1 & 0.01 & 0.999911 & 0.638787 & 0.998910 & 0.627267 \\
 &  & 0.01 & 0.1 & 0.954438 & 0.887484 & 0.953438 & 0.885983 \\
2 & 0 & 0.1 & 0.01 & 0.999913 & 0.712338 & 0.998912 & 0.702947 \\
 &  & 0.01 & 0.1 & 0.954440 & 0.884025 & 0.953440 & 0.882563 \\
 & 1 & 0.1 & 0.01 & 0.999622 & 0.725851 & 0.998622 & 0.716879 \\
 &  & 0.01 & 0.1 & 0.917015 & 0.852051 & 0.916015 & 0.850766 \\
 & 2 & 0.1 & 0.01 & 0.999200 & 0.748196 & 0.998199 & 0.739937 \\
 &  & 0.01 & 0.1 & 0.866525 & 0.801327 & 0.865525 & 0.800141 \\
10 & 0 & 0.1 & 0.01 & 0.994432 & 0.888765 & 0.993432 & 0.885794 \\
 & 5 & 0.1 & 0.01 & 0.987613 & 0.897157 & 0.986613 & 0.894680 \\
 & 10 & 0.1 & 0.01 & 0.978199 & 0.900957 & 0.977199 & 0.898987 \\
\end{tabular}
\end{ruledtabular}
\end{table}

\newpage

\begin{figure}
\centering
\includegraphics[height=5in, width=6.5in, angle=0]{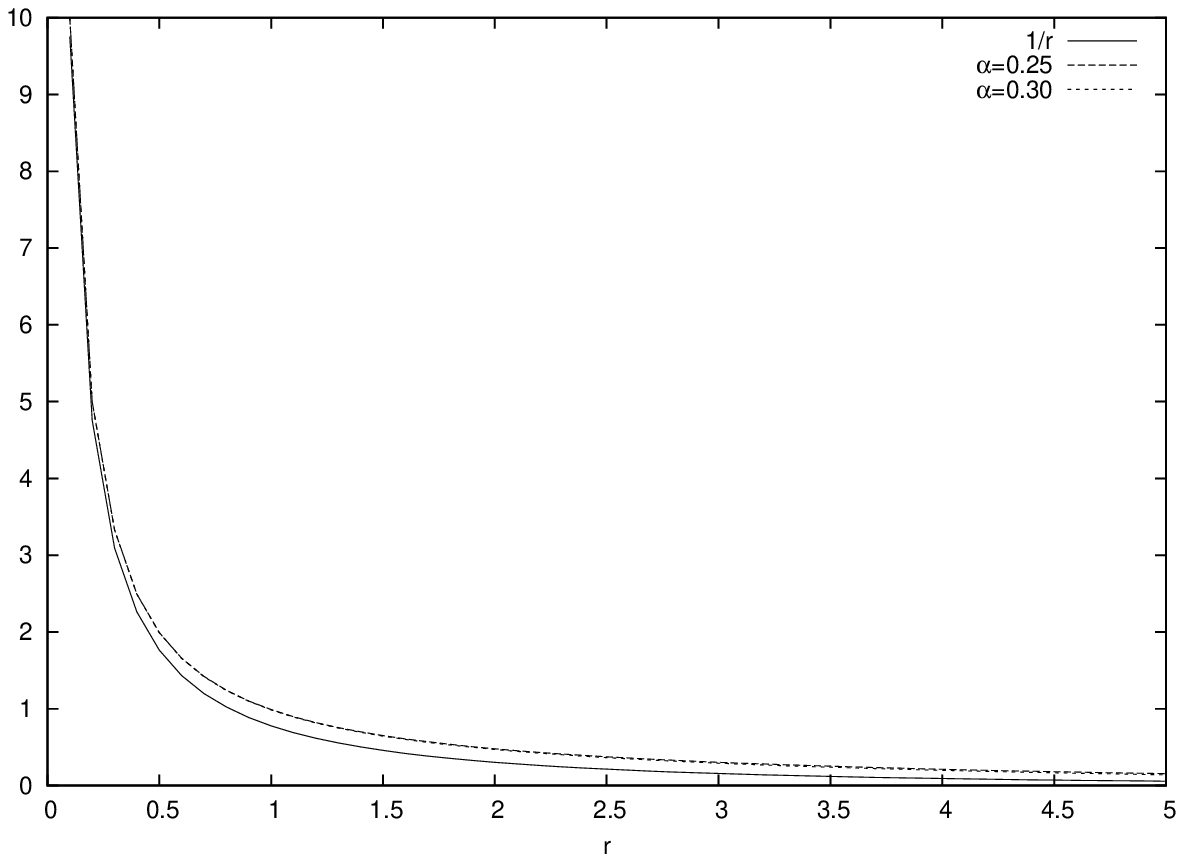}
\caption{Comparison of $1/r$ (full line) with $2\alpha e^{-\alpha
r }(1-e^{-\alpha r })$ for different values of $\alpha=0.25,
0.30$, respectively.}
\end{figure}


\newpage



\begin{thebibliography}{99}

\bibitem{ref1} L.~Z.~Yi, Y.~F.~Diao, J.~Y.~Liu, and C.~S.~Jia, Phys. Lett. A {\bf 333}, 212 (2004).


\bibitem{ref2} Y.~F.~Diao, L.~Z.~Yi, and C.~S.~Jia, Phys. Lett. A {\bf 332}, 157 (2004).


\bibitem{ref3} S.~Haouat, and L.~Chetouani, Phys. Scr. {\bf 77}, 025005 (2008).


\bibitem{ref4} H.~Yukawa, Proc. Phys. Math. Soc. Jpn. {\bf 17}, 48 (1935).


\bibitem{ref5} A.~D.~Alhaidari, H.~Bahlouli, and M.~S.~Abdelmonem, J. Phys. A {\bf 41}, 032001 (2008).

\bibitem{ref6} R.~Sever, and C.~Tezcan, Phys. Rev. A {\bf 41}, 5205 (1990).


\bibitem{ref7} R.~Sever, and C.~Tezcan, Phys. Rev. A {\bf 37}, 3158 (1988).


\bibitem{ref8} R.~Sever, and C.~Tezcan, Phys. Rev. A {\bf 35}, 2725 (1987).


\bibitem{ref9} R.~Sever, and C.~Tezcan, Phys. Rev. A {\bf 36}, 1045 (1987).


\bibitem{ref10} O.~von Roos, Phys. Rev. B {\bf 27}, 7547 (1983).



\bibitem{ref11} J.~C.~Slater, Phys. Rev. {\bf 76}, 1592 (1949).



\bibitem{ref12} L.~Serra, and E.~Lipparini, Europhys. Lett. {\bf 40}, 667 (1997).


\bibitem{ref13} T.~Gora, and F.~Williams, Phys. Rev. {\bf 177}, 1179 (1969).


\bibitem{ref14} C.~S.~Jia, J.~Y.~Liu, P.~Q.~Wang, and C.~S.~Che, Phys. Lett. A {\bf 369},
274 (2007).

\bibitem{ref15} C.~S.~Jia, and A.~S.~Dutra, Ann. Phys. (N.Y.) {\bf 323}, 566 (2008).

\bibitem{ref16} A.~D.~Alhaidari, Phys. Lett. A {\bf 322}, 72 (2004).

\bibitem{ref17} M.~M.~Panja, R.~Dutt, and Y.~P.~Varshini, Phys. Rev. A {\bf 42}, 106 (1990).

\bibitem{ref18} R.~L.~Green, and C.~Aldrich, Phys. Rev. A {\bf 14}, 2363 (1976).

\bibitem{ref19} W.~C.~Qiang, and S.~H.~Dong, Phys. Lett. A {\bf 368}, 13 (2007).


\bibitem{ref20} A.~D.~Alhaidari, Found. Phys. {\bf 40}, 1088 (2010).


\bibitem{ref21} M.~Abramowitz, and I.~A.~Stegun, (eds.),
\textit{Handbook of Mathematical Functions with Formulas, Graphs,
and Mathematical Tables} (New York, 1965).

\bibitem{ref22} E.~Z.~Liverts, and V.~B.~Mandelzweig, Ann. Phys. {\bf 324}, 388 (2009).

\bibitem{ref23} R.~L.~Hall, Phys. Lett. A {\bf 372}, 12 (2007).

\bibitem{ref24} A.~D.~Alhaidari, H.~Bahlouli, and A.~Al-Hasan, Phys. Lett. A {\bf 349}, 87 (2006).

\bibitem{ref25} T.~Barakat, M.~Odeh, and O.~Mustafa, J. Phys. A {\bf 31}, 3469 (1998).

\bibitem{ref26} S.~Ikhdair, Eur. Phys. J. A {\bf 40}, 143 (2009).


\bibitem{ref27} T.~Q.~Dai, and Y.~F.~Cheng, Phys. Scr. {\bf 79},
015007 (2009).
\end{thebibliography}
\end{document}